\documentclass[twocolumn,prb]{revtex4}
\usepackage{amsfonts}
\usepackage[T1]{fontenc}
\usepackage{amsmath,amsbsy,amssymb,graphicx}
\usepackage{times}

\begin{document}

\title{Nonlinearity-induced transition \\
in nonlinear Su-Schrieffer-Heeger model and nonlinear higher-order
topological system}
\author{Motohiko Ezawa}
\affiliation{Department of Applied Physics, University of Tokyo, Hongo 7-3-1, 113-8656,
Japan}

\begin{abstract}
We study the topological physics in nonlinear Schr\"{o}dinger systems on
lattices. We employ the quench dynamics to explore the phase diagram, where
a pulse is given to a lattice point and we analyze its time evolution. There
are two system parameters $\lambda $ and $\xi $, where $\lambda $ controls
the hoppings between the neighboring links and $\xi $ controls the
nonlinearity. The dynamics crucially depends on these system parameters.
Based on analytical and numerical studies, we derive the phase diagram of
the nonlinear Su-Schrieffer-Heeger (SSH) model in the ($\lambda ,\xi $)
plane. It consists of four phases. The topological and trivial phases emerge
when the nonlinearity $\xi $ is small. The nonlinearity-induced localization
phase emerges when $\xi $ is large. We also find a dimer phase as a result
of a cooperation between the hopping and nonlinear terms. A similar analysis
is made of the nonlinear second-order topological system on the breathing
Kagome lattice, where a trimer phase appears instead of the dimer phase.
\end{abstract}

\maketitle

\section{Introduction}

Topological phases have attracted much attention in the context of solid
state materials\cite{Hasan,Qi} with the emergence of topological edge
states. They are generalized to higher-order topological phases\cite%
{Fan,Science,APS,Peng,Lang,Song,Bena,Schin,FuRot,EzawaKagome,Khalaf}, where
topological corner states and topological hinge states emerge. Recently,
they are also found in various linear systems such as photonic\cite%
{KhaniPhoto,Hafe2,Hafezi,WuHu,TopoPhoto,Ozawa16,Ley,KhaniSh,Zhou,Jean,Ota18,Ozawa,Ota19,OzawaR,Hassan,Ota,Li,Yoshimi,Kim,Iwamoto21}%
, acoustic\cite{Prodan,TopoAco,Berto,Xiao,He,Abba,Xue,Ni,Wei,Xue2},
mechanical\cite%
{Lubensky,Chen,Nash,Paul,Sus,Sss,Huber,Mee,Kariyado,Hannay,Po,Rock,Takahashi,Mat,Taka,Ghatak,Wakao}
and electric circuit\cite%
{TECNature,ComPhys,Hel,Lu,YLi,EzawaTEC,Research,Zhao,EzawaLCR,EzawaSkin,Garcia,Hofmann,EzawaMajo,Tjunc,Lee,Kot}
systems. Now, nonlinear topological photonics is an emerging field\cite%
{Ley,Zhou,Smi,Kruk,MacZ}, where nonlinearity is naturally introduced by the
Kerr effect. Nonlinear higher-order topological phases have been
experimentally studied in photonics\cite{Zange,Kirch}. Topological edge
states and topological corner states have been observed in nonlinear systems
just as in linear systems.

It is a hard task to construct a general theory of the topological physics
in nonlinear systems because there are many ways to introduce nonlinearity.
It would be necessary to make individual studies of typical nonlinear models
to achieve at a systematic understanding. We studied the dimerized
Toda-lattice model\cite{TopoToda} and a nonlinear mechanical system\cite%
{MechaRot} in previous works. These models contain the Su-Schrieffer-Heeger
(SSH) model as an essential term. Indeed, these models are reduced to the
dynamical SSH model provided the nonlinear term is ignored, where the
topological number is well defined and the zero-mode edge state emerges in
the topological phase. Then, we carried out numerical analysis to show the
topological physics is valid even in the presence of the nonlinear term.\
These models have only two phases, the topological phase and the trivial
phase in the phase diagram in the ($\lambda ,\xi $) plane, with $\lambda $
the\ dimerization parameter and $\xi $ the nonlinearity parameter.

\begin{figure}[t]
\centerline{\includegraphics[width=0.48\textwidth]{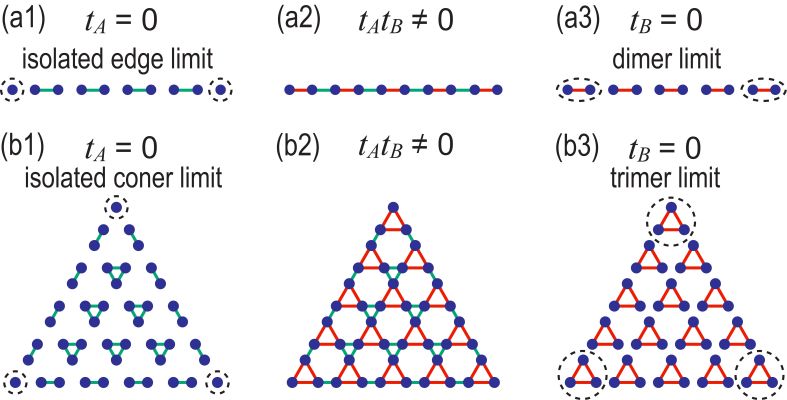}}
\caption{Illustration of (a) a dimerized lattice and (b) a breathing Kagome
lattice with (a) $t_{A}=0$, (b) $t_{A}t_{B}\neq 0$ and (c) $t_{B}=0$. A line
(triangle) contains many small segments (triangles). At the edges (corners)
of the chain (triangle), there are two (three) isolated atoms for $t_{A}=0$,
while there are dimer (trimer) states for $t_{B}=0$. They are marked by
dotted circles. The size of the line (triangle) is $L=5$. }
\label{FigKagomeIllust}
\end{figure}

In this paper, we study the quench dynamics governed by a nonlinear Schr\"{o}%
dinger equation consisting of the hopping term with the hopping matrix $%
M_{nm}$ and the nonlinear term proportional to the nonlinearity parameter $%
\xi $. In the quench dynamics, we give a pulse to a lattice point and
explore its time evolution. The dynamics is sensitive to the presence of the
topological edge and corner states. We perform a numerical analysis in a
wide region of parameters and construct phase diagrams in the ($\lambda ,\xi 
$) plane. We are interested in the systems which describe nontrivial
topological dynamics in the linear limit ($\xi =0$). As explicit examples,
we take $M_{nm}$ on the SSH lattice and on the breathing Kagome lattice. We
confirm analytically the validity of the topological dynamics in the weak\
nonlinearity regime ($\xi \ll 1$) based on the first-order perturbation
theory in $\xi $. We show that the topological phase boundary between the
topological and trivial phases is well defined and\ not modified in this
weak nonlinearity regime.\ In the strong nonlinearity regime ($\xi \gg 1$)
where the nonlinear term is dominant, we obtain analytically the
nonlinearity-induced localization phase, where the state is localized due to
the nonlinear term. It is unrelated to the topological physics because the
term $M_{nm}$ is irrelevant in this regime. The transition from the weak to
the strong nonlinearity regime is a transition from extended states to
localized states. We have also found a new phase formed by a cooperative
effect of these two terms, which is the oscillation-mode phase in the
vicinity of the dimerized nonlinear SSH model and the trimerized breathing
Kagome model illustrated in Fig.\ref{FigKagomeIllust}(a3) and (b3),
respectively.

This paper is composed as follows. In Section \ref{SecSelfTrap}, we review
the nonlinear Schr\"{o}dinger equation. We discuss it analytically in the
linear limit ($\xi =0$), in the weak nonlinearity regime ($\xi \ll 1$) and
in the strong nonlinearity regime ($\xi \gg 1$). We find that the
topological phase transition point does not change in the weak nonlinear
regime. On the other hand, the system turns into\ the nonlinearity-induced
localization phase in the strong nonlinearity regime.\ In Section \ref%
{SecSSH}, we explicitly study the nonlinear SSH model, where the phase
diagram is determined by a numerical analysis. It consists of the
topological phase, the trivial phase, the nonlinearity-induced localization
phase and the dimer phase. We discuss the origin of the dimer phase as an
operative effect of the hopping term and the nonlinear term. In Section \ref%
{SecKagome} we explicitly study the nonlinear second-order topological phase
on the breathing Kagome lattice, where the phase diagram is constructed by a
numerical analysis. The analysis and the results are quite similar to those
in the nonlinear SSH model except for the trimer phase replacing the dimer
phase.

\section{Nonlinear Schr\"{o}dinger equation\label{SecSelfTrap}}

A typical nonlinear equation is the nonlinear Schr\"{o}dinger equation,%
\begin{equation}
i\frac{\partial \psi }{\partial t}+\varepsilon \frac{\partial ^{2}\psi }{%
\partial x^{2}}+\xi \left\vert \psi \right\vert ^{2}\psi =0,
\end{equation}%
where the third term is a nonlinear term. It is introduced by the Kerr
effect in the case of photonic systems\cite{Szameit,Chris}. The nonlinearity
is controlled by the parameter $\xi $, where large $\xi $ indicates strong
nonlinearity. 
There is a lattice version of the above equation, 
\begin{equation}
i\frac{d\psi _{n}}{dt}+\varepsilon \left( \psi _{n+1}-2\psi _{n}+\psi
_{n-1}\right) +\xi \left\vert \psi _{n}\right\vert ^{2}\psi _{n}=0,
\label{DNLS}
\end{equation}%
which is called the discrete nonlinear Schr\"{o}dinger equation\cite{Cai,Kev}%
. There are two conserved quantities. One is the Hamiltonian\cite%
{Eil,Szameit,Korab},%
\begin{equation}
H=\sum_{n=1}^{N}\left( \varepsilon \left\vert \psi _{n+1}-\psi
_{n}\right\vert ^{2}-\frac{\xi }{2}\left\vert \psi _{n}\right\vert
^{4}\right) .
\end{equation}%
and the other is the excitation number,%
\begin{equation}
N_{\text{exc}}=\sum_{n=1}^{N}\left\vert \psi _{n}\right\vert ^{2}.
\label{Normal}
\end{equation}

The discrete nonlinear Schr\"{o}dinger equation (\ref{DNLS}) is defined on
the one-dimensional lattice. It is generalized to a nonlinear equation on an
arbitrary lattice\cite{Eil,Szameit,Chris},%
\begin{equation}
i\frac{d\psi _{n}}{dt}+\sum_{m=1}^{N}M_{nm}\psi _{m}+\xi \left\vert \psi
_{n}\right\vert ^{2}\psi _{n}=0,  \label{DST}
\end{equation}%
where $M_{nm}$ represents a hopping matrix, and $N$ is the number of the
lattice sites. 
We investigate such
a system that contains the topological and trivial phases provided the
nonlinear term is ignored. We study analytically and numerically the phase
diagram of the model (\ref{DST}). The main issue is how the topological
phase defined in the linear model ($\xi =0$) is robust against the
introduction of the nonlinear term.

There are two conserved quantities\cite{Korab}. One is the Hamiltonian%
\begin{equation}
H=\sum_{n=1}^{N}\left( -M_{nm}\psi _{n}^{\ast }\psi _{m}-\frac{\xi }{2}%
\left\vert \psi _{n}\right\vert ^{4}\right) ,
\end{equation}%
and the other is the excitation number (\ref{Normal}).

We analyze the quench dynamics by imposing an initial condition%
\begin{equation}
\psi _{n}\left( t\right) =\delta _{n,m}\qquad \text{at}\qquad t=0.
\label{IniCon}
\end{equation}%
Namely, giving an delta-function type input at the site $m$ initially, we
study its time evolution. Because of the conservation rule (\ref{Normal}),
the condition%
\begin{equation}
\sum_{n=1}^{N}\left\vert \psi _{n}\right\vert ^{2}=1  \label{NormConser}
\end{equation}%
is imposed throughout the time evolution.

A comment is in order. It is possible to eliminate the nonlinearity
parameter $\xi $ entirely from Eq.(\ref{DST}). By setting $\psi _{j}=\psi
_{j}^{\prime }/\sqrt{\xi }$, we may rewrite (\ref{DST}) as%
\begin{equation}
i\frac{d\psi _{n}^{\prime }}{dt}+\sum_{m}M_{nm}\psi _{m}^{\prime
}+\left\vert \psi _{n}^{\prime }\right\vert ^{2}\psi _{n}^{\prime }=0.
\label{DST1}
\end{equation}%
The initial condition (\ref{IniCon}) is replaced by 
\begin{equation}
\psi _{n}^{\prime }\left( t=0\right) =\sqrt{\xi }\delta _{n,m}.
\label{IniCon22}
\end{equation}%
Namely, the quench dynamics subject to Eq.(\ref{DST}) is reproduced by the
nonlinear equation (\ref{DST1}) with the modified initial condition (\ref%
{IniCon22}). Consequently, it is possible to use a single sample to
investigate the quench dynamics at various nonlinearity $\xi $ only by
changing the initial condition as in (\ref{IniCon22}). Nevertheless, we use
the form of Eq.(\ref{DST}) throughout the paper to make the nonlinear effect
manifest.

\begin{figure*}[t]
\centerline{\includegraphics[width=0.98\textwidth]{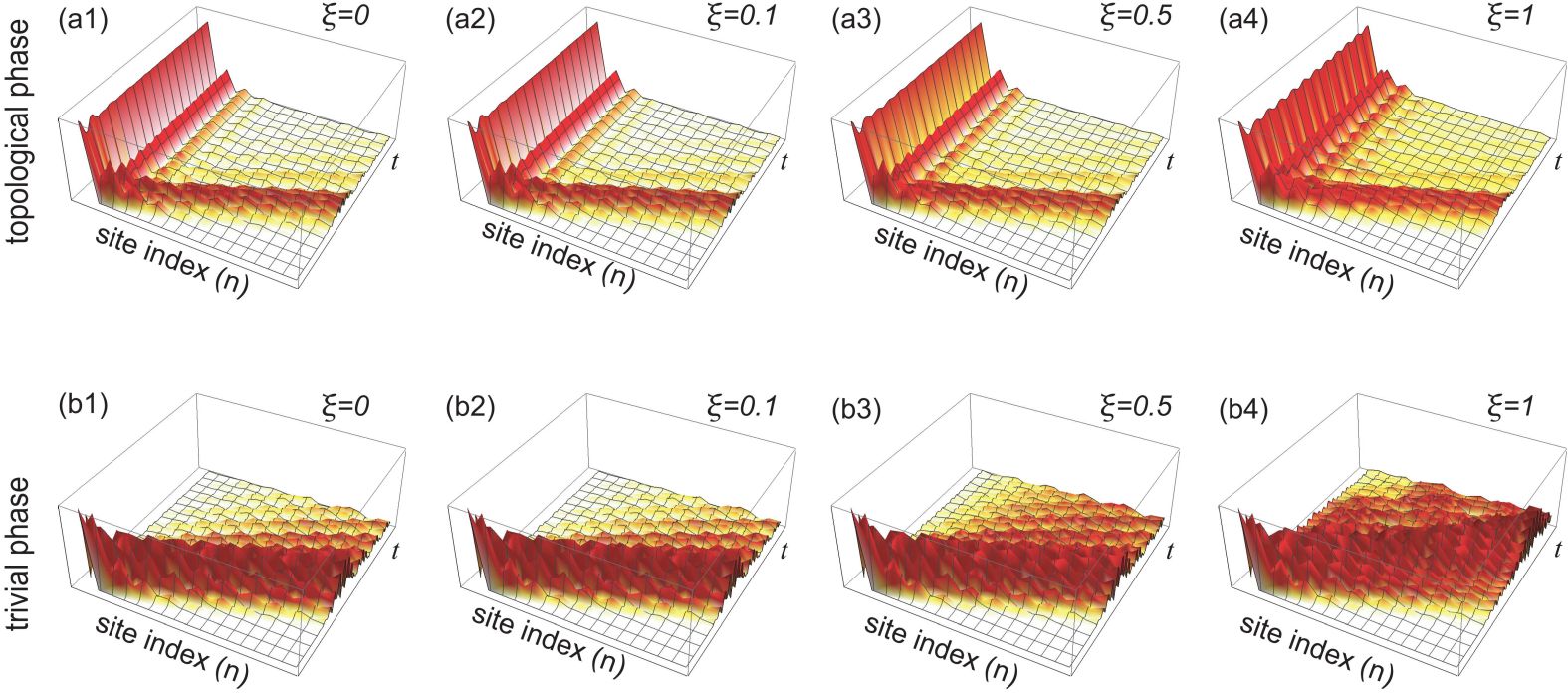}}
\caption{Bird's eye's view of time evolution of the amplitude $|\protect\psi%
_{n}|$\ in the nonlinear SSH model. The horizontal axes are the site index $%
n $ and the time $t$ ranging $0\leq t \leq 30$. (a1)$\sim$(a4) Topological
phase ($\protect\lambda =-0.5$). (b1)$\sim $(b4) Trivial phase ($\protect%
\lambda =0.5$). We have set $\protect\xi =0$ for (a1) and (b1), $\protect\xi %
=0.1$ for (a2) and (b2), $\protect\xi =0.5$ for (a3) and (b3), and $\protect%
\xi =1$ for (a4) and (b4). }
\label{FigDynamics}
\end{figure*}

\subsection{Linearized model}

We first study the linear limit by setting $\xi =0$,%
\begin{equation}
i\frac{d\psi _{n}}{dt}+\sum_{m}M_{nm}\psi _{m}=0.  \label{TopoDyna}
\end{equation}%
We diagonalize $M_{nm}$ as 
\begin{equation}
M\bar{\psi}_{p}=E_{p}\bar{\psi}_{p},  \label{EigenA}
\end{equation}%
where $p$ labels the eigen index, $1\leq p\leq N$. Then, we obtain decoupled
equations%
\begin{equation}
i\frac{d\bar{\psi}_{p}}{dt}+E_{p}\bar{\psi}_{p}=0,
\end{equation}%
whose solutions are given by%
\begin{equation}
\bar{\psi}_{p}\left( t\right) =\exp \left[ -itE_{p}\right] \bar{\psi}%
_{p}\left( 0\right) .  \label{LSol}
\end{equation}%
The initial state is expanded as 
\begin{equation}
\psi _{n}\left( 0\right) =\delta _{n,m}=\sum_{p}c_{p}\bar{\psi}_{p}\left(
0\right) .  \label{Expand}
\end{equation}%
Because Eq.(\ref{TopoDyna}) is a linear model, the topological numbers
defined with respect to $M_{nm}$ determines the topological phases of the
system.

There are localized states in a topological phase known as zero-mode edge
states in the one-dimensional topological phase and zero-mode corner states
in the\ two-dimensional second-order topological phase. We impose the
initial condition (\ref{IniCon}) with $m=1$, or%
\begin{equation}
\psi _{n}\left( 0\right) =\delta _{n,1},  \label{IniConA}
\end{equation}%
where the site $n=1$ denotes the left edge of a chain or the top corner of a
triangle. The zero-mode edge (corner) state is given in terms of an
eigenstate of $M_{nm}$, which we assume to be $\bar{\psi}_{1}$ with $E_{1}=0$
in (\ref{EigenA}). With the use of the expansion (\ref{Expand}), the
zero-mode edge (corner) state $\bar{\psi}_{1}$ is well approximated by $\psi
_{1}$ at $t=0$, or%
\begin{equation}
\psi _{1}\left( 0\right) \simeq c_{1}\bar{\psi}_{1}\left( 0\right) .
\end{equation}%
Since the zero-mode edge (corner) state has the zero energy, there is no
dynamics,%
\begin{equation}
\psi _{1}\left( t\right) =c_{1}\bar{\psi}_{1}\left( 0\right) .
\end{equation}%
As a result, there remains a finite component $c_{1}$ at the edge (corner)
site even after time evolution.

On the other hand, there is no zero-mode localized state at the edge
(corner) in the trivial phase, and the state $\psi _{n}\left( t\right) $
rapidly penetrates into the bulk. Consequently, it is possible to
differentiate the topological and trivial phases numerically by checking
whether there remains a finite component or not under the initial condition (\ref{IniConA}).

\begin{figure*}[t]
\centerline{\includegraphics[width=0.98\textwidth]{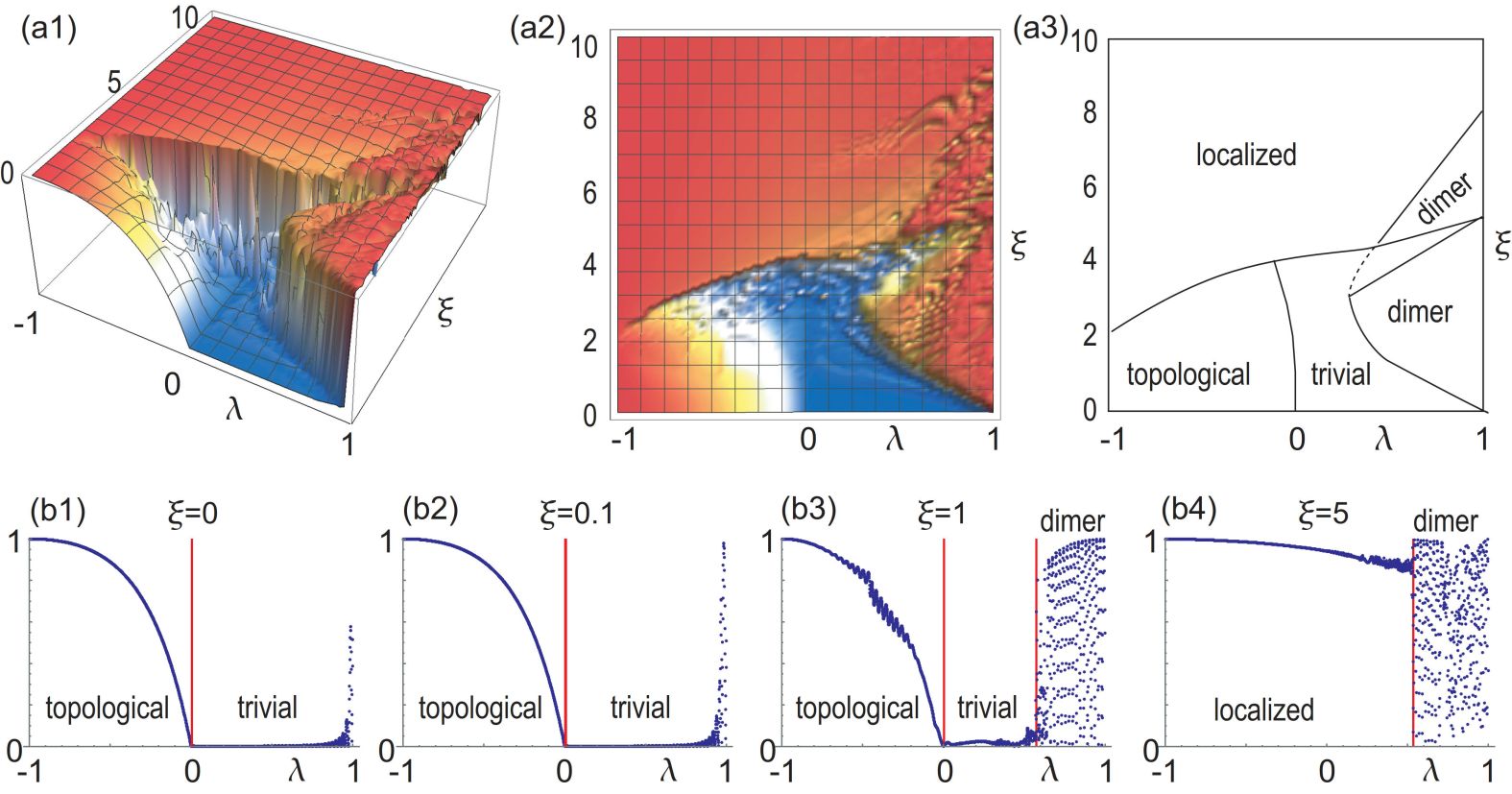}}
\caption{(a1)$\sim$(a3) Phase diagram of the nonlinear SSH model. (a1)
Bird's eye's view, (a2) top view and (a3) schematic illustration of the
phase diagram. (b1)$\sim$(b4) Amplitude $|\protect\psi_1|$ as a function of 
$\protect\lambda $ for various $\protect\xi $. (b1) $\protect\xi =0$, (b2) 
$\protect\xi =0.1$, (b3) $\protect\xi =1$, and (b4) $\protect\xi =5$.}
\label{FigSSHDiagram}
\end{figure*}

\subsection{Weak nonlinear regime}

\label{SecWeak}

We study the weak nonlinear regime of Eq.(\ref{DST}), which is the regime
where the first-order perturbation in $\xi $ is valid. We may insert the
linear solution (\ref{LSol}) to the nonlinear term proportional to $\xi $ in Eq.(\ref{DST}), i.e.,%
\begin{equation}
\xi \left\vert \psi _{n}(t)\right\vert ^{2}\psi _{n}(t)=\xi \left\vert \psi
_{n}\left( 0\right) \right\vert ^{2}\psi _{n}(t)+O(\xi ^{2}),
\end{equation}%
and obtain%
\begin{equation}
i\frac{d\psi _{n}}{dt}+\sum_{m}\overline{M}_{nm}\psi _{m}=0,
\label{TopoDynaA}
\end{equation}%
where%
\begin{equation}
\overline{M}_{nm}\equiv M_{nm}+\delta _{nm}\xi \left\vert \psi _{n}\left(
0\right) \right\vert ^{2}.
\end{equation}%
The second term $\delta _{nm}\xi \left\vert \psi _{n}\left( 0\right)
\right\vert ^{2}$ may be regarded as an on-site random potential in the
linearized model. Then, the topological phase is robust against the on-site
potential as far as the bulk gap does not close or equivalently $\xi
\left\vert \psi _{n}\left( 0\right) \right\vert ^{2}$ is smaller than the
gap of $M_{nm}$. Consequently, in the weak nonlinearity regime, the
topological number is well defined, and the topological phase boundary is
unchanged as $\xi $\ increases. See explicit examples in Sec.\ref{SecSSH}
and Sec.\ref{SecKagome}, where the topological phase diagrams are
numerically constructed.

\subsection{Strong nonlinear regime}

\label{SecStrong}

We next study the strong nonlinear regime ($\xi \gg 1$), which is the regime
where the hopping term is negligible with respect to the nonlinear term. We
may approximate Eq.(\ref{DST}) as%
\begin{equation}
i\frac{d\psi _{n}}{dt}=-\xi \left\vert \psi _{n}\right\vert ^{2}\psi _{n},
\end{equation}%
where all equations are separated. We set%
\begin{equation}
\psi _{n}\left( t\right) =r_{n}e^{i\theta _{n}\left( t\right) },  \label{rt}
\end{equation}%
and make an ansatz that $r_{n}$ is a constant in the time $t$. This ansatz
is confirmed numerically in Sec.\ref{SecSSH} and Sec.\ref{SecKagome}. Then,
the solution is given by%
\begin{equation}
\theta _{n}=\xi r_{n}^{2}t+c.
\end{equation}%
Hence, the amplitude does not decrease. Due to the norm conservation (\ref{NormConser}), 
we find $\psi _{m}\left( t\right) =\delta _{nm}.$\ Namely,
the state $\psi _{n}$\ does not spread under the initial condition (\ref{IniCon}), 
as we will see by taking an explicit model in Sec.\ref{SecDNS}.
This phase may be referred to as the nonlinearity-induced localization phase.

We note that there is no concept of topology in the strong nonlinear regime
because the $M_{nm}$ term is irrelevant. This property is confirmed in Sec.\ref{SecSSH} and Sec.\ref{SecKagome} based on explicit examples.

\subsection{Dynamics of edge or corner state}

We consider the case where the edge (corner) is perfectly decoupled from\
all sites in the bulk. See Fig.\ref{FigKagomeIllust}(a1) and (b1) for
examples. It is enough to solve the single differential equation, 
\begin{equation}
i\frac{d\psi _{1}}{dt}=\varepsilon \psi _{1}-\xi \left\vert \psi
_{1}\right\vert ^{2}\psi _{1},
\end{equation}%
where $\varepsilon $\ is the on-site energy of the site $n=1$. As in the
strong nonlinear regime, we assume the condition (\ref{rt}), and we obtain 
\begin{equation}
-r_{n}e^{i\theta _{n}\left( t\right) }\frac{d\theta _{n}\left( t\right) }{dt}%
=\varepsilon r_{n}e^{i\theta _{n}\left( t\right) }-\xi r_{n}^{3}e^{i\theta
_{n}\left( t\right) },
\end{equation}%
or 
\begin{equation}
\frac{d\theta _{n}\left( t\right) }{dt}=-\varepsilon +\xi r_{n}^{2}.
\end{equation}%
The solution is given by%
\begin{equation}
\theta _{n}=\left( -\varepsilon +\xi r_{n}^{2}\right) t+c,
\end{equation}%
with a constant $c$. It shows that the amplitude does not change as a
function of the time $t$.

\section{Nonlinear SSH model\label{SecSSH}}

\subsection{Model}

We consider explicit models. The first example is the nonlinear SSH model%
\cite{Hadad,Gor,Tulo,Zhou}, where $M_{nm}$ is given by%
\begin{equation}
M_{nm}=-\delta _{nm}\left( t_{A}+t_{B}\right) +\left( t_{A}\delta
_{n,2m-1}+t_{B}\delta _{2n,2m+1}\right)   \label{HoppingMmn}
\end{equation}%
We illustrate the lattice model of the SSH model in Fig.\ref{FigKagomeIllust}%
, which is a dimerized lattice. For $t_{A}=0$, two edge sites are perfectly
decoupled whereas all other bulk sites are dimerized as in Fig.\ref%
{FigKagomeIllust}(a1). On the other hand, for $t_{B}=0$, all of the sites
are dimerized as in Fig.\ref{FigKagomeIllust}(a3).

The equations of motion (\ref{DST}) read%
\begin{align}
i\frac{d\psi _{2n-1}}{dt}& =t_{B}\left( \psi _{2n-2}-\psi _{2n-1}\right)
+t_{A}\left( \psi _{2n}-\psi _{2n-1}\right)  \notag \\
& -\xi \left\vert \psi _{2n-1}\right\vert ^{2}\psi _{2n-1},  \label{EqA} \\
i\frac{d\psi _{2n}}{dt}& =t_{A}\left( \psi _{2n-1}-\psi _{2n}\right)
+t_{B}\left( \psi _{2n+1}-\psi _{2n}\right)  \notag \\
& -\xi \left\vert \psi _{2n}\right\vert ^{2}\psi _{2n},  \label{EqB}
\end{align}%
with alternating bondings $t_{A}$ and $t_{B}$. We introduce the dimerization
control parameter defined by%
\begin{equation}
\lambda =\frac{t_{A}-t_{A}}{t_{A}+t_{A}},  \label{SpringCon}
\end{equation}%
where $|\lambda |\leq 1$.

\begin{figure}[t]
\centerline{\includegraphics[width=0.48\textwidth]{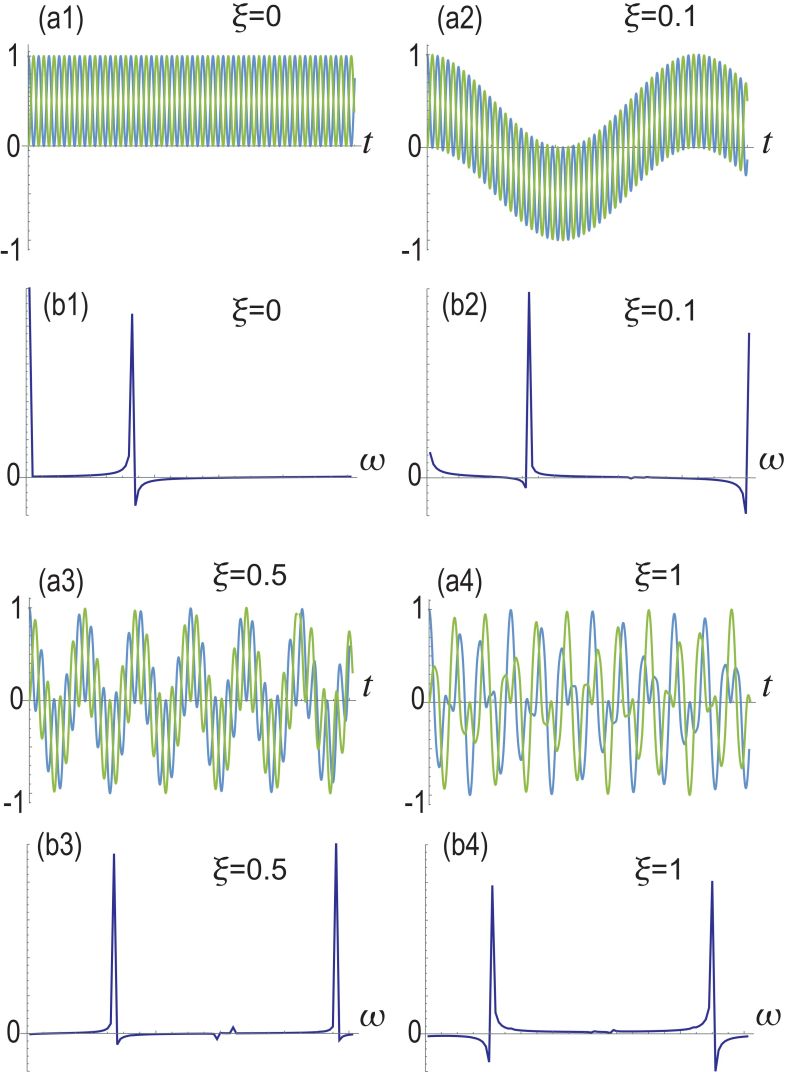}}
\caption{(a1)$\sim$(a4) Time evolution of Re[$\protect\psi_1(t)$] in the
nonlinear Schr\"{o}dinger model on the dimer described by Eqs.(\protect\ref{dimer1}) and (\protect\ref{dimer2}). 
(a1), (b1) $\protect\xi =0$; (a2),
(b2) $\protect\xi =0.1$; (a3), (b3) $\protect\xi =0.5$; (a4), (b4) $\protect%
\xi =1$. (a1)$\sim$(a4) The vertical axis is Re[$\protect\psi_1(t)$] and the
horizontal axis is the time $t$. (b1)$\sim$(b4) Fourier component of Re[$%
\protect\psi_1(\protect\omega )$]. The horizontal axis is the frequency $%
\protect\omega$, which is the Fourier component of the time $t$, while the
vertical axis is Re[$\protect\psi_1(\protect\omega )$].}
\label{FigDimerDynamics}
\end{figure}

\begin{figure*}[t]
\centerline{\includegraphics[width=0.98\textwidth]{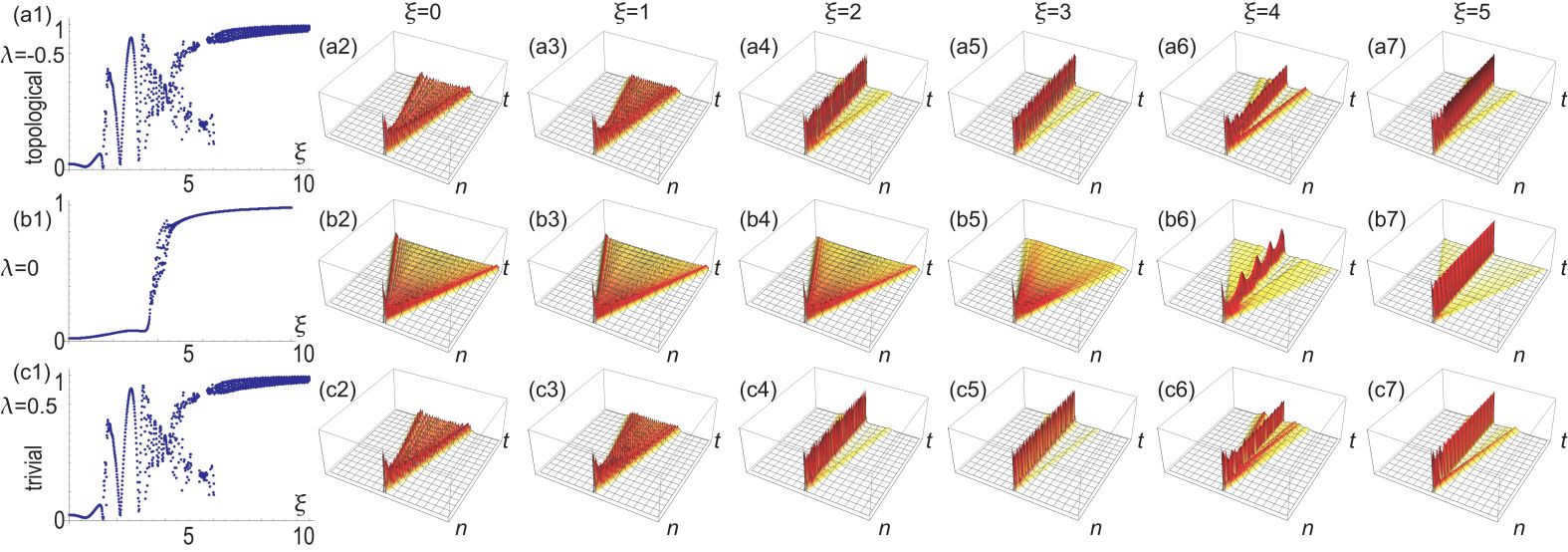}}
\caption{ (a1)$\sim$(c1) Amplitude $|\protect\psi _{1}|$ after enough time
as a function of $\protect\xi$. (a2)$\sim $(a7), (b2)$\sim $(b7) and (c2)$%
\sim $(c7) Time evolution of $|\protect\psi _{n}|$ in the discrete nonlinear
Schr\"{o}dinger equation (\protect\ref{DNLS}). (a1)$\sim$(a7) topological
phase at $\protect\lambda =-0.5$, (a1)$\sim$(a7) topological phase boundary
at $\protect\lambda =0$, and (a1)$\sim$(a7) trivial phase at $\protect%
\lambda =0.5$.}
\label{FigLocalized}
\end{figure*}

\subsection{Phase diagram}

Starting from the initial condition (\ref{IniConA}), we explore the time
evolution of $\psi _{n}$ for various $\xi $\ and show the results in Fig.\ref%
{FigDynamics}(b). As indicated by a general consideration given before, we
find that there remains a finite component at the edge site ($n=1$) in the
topological phase, while it is almost zero in the trivial phase.

We show the absolute value of $\psi _{1}$ after enough time as a function of 
$\lambda $ for various $\xi $\ in Fig.\ref{FigSSHDiagram}(b1)$\sim $(b4).
First, we study the linear model as shown in Fig.\ref{FigSSHDiagram}(b1).
The amplitude $\left\vert \psi _{1}\right\vert $\ is finite in the
topological phase, while it is almost zero in the trivial phase. The overall
structure is almost identical in the linear limit ($\xi =0$) and in the weak
nonlinear regime ($\xi =0.1$)\ as shown in Fig.\ref{FigSSHDiagram}(b2). For
medium nonlinearity ($\xi =1$), there appears an oscillation mode for $%
\lambda \geq 0.55$, as shown in Fig.\ref{FigSSHDiagram}(b3). We will argue
that this is due to the dimerization effect in Sec.\ref{SecDimer}. For
strong nonlinearity ($\xi =5$), the amplitude $\left\vert \psi
_{1}\right\vert $\ is almost 1 for $\lambda \leq 0.52$. We have already
argued that this is due to the nonlinearity-induced localization in Sec.\ref%
{SecStrong}.

There are four phases. First, we have the topological and trivial phases in
the weak nonlinear regime. The topological phase boundary is almost
independent of the nonlinearity $\xi $ . The amplitude gradually decreases
from $1$ to $0$ depending on the dimerization from $\lambda =-1$ to $\lambda
=0$, as we have argued in Sec.\ref{SecWeak}. On the other hand, there is a
nonlinearity-induced localization phase for large $\xi $. The amplitude is
almost $1$ entirely in the nonlinearity-induced localization phase, as we
have argued in Sec.\ref{SecStrong}.

In addition, there is a dimer phase in the vicinity of $\lambda \simeq 1$,
where the system is almost dimerized. The states $\psi _{1}$ and $\psi _{2}$%
\ oscillate between the two adjacent sites ($n=1,2$) at the edge.
Furthermore, the trivial phase penetrates into the dimer phase for $\lambda
\geq 0.25$ in Fig.\ref{FigSSHDiagram}(a2) as in (a3).

\subsection{Topological number}

The hopping matrix (\ref{HoppingMmn}) leads to the SSH Hamiltonian in the
momentum space,%
\begin{equation}
M\left( k\right) =-\left( t_{A}+t_{B}\right) I_{2}+\left( 
\begin{array}{cc}
0 & t_{A}+t_{B}e^{-ik} \\ 
t_{A}+t_{B}e^{ik} & 0%
\end{array}%
\right) .  \label{EqK}
\end{equation}%
The topological number is the Berry phase defined by%
\begin{equation}
p_{i}=\frac{1}{2\pi }\int_{9}^{2\pi }A\left( k\right) dk,
\label{ChiralIndex}
\end{equation}%
where $A\left( k\right) =-i\left\langle \psi (k)\right\vert \partial
_{k}\left\vert \psi (k)\right\rangle $ is the Berry connection with $\psi
(k) $ the eigenfunction of $M\left( k\right) $. We obtain $\Gamma =1$ for $%
\lambda <0$ and $\Gamma =0$ for $\lambda >0$. It is known that the SSH
system is topological for $\lambda <0$ and trivial for $\lambda >0$. There
are two isolated edge states in the limit $\lambda \simeq -1$, while all of
the states are dimerized in the limit $\lambda \simeq 1$: See Fig.\ref%
{FigKagomeIllust}(a).

\begin{figure*}[t]
\centerline{\includegraphics[width=0.98\textwidth]{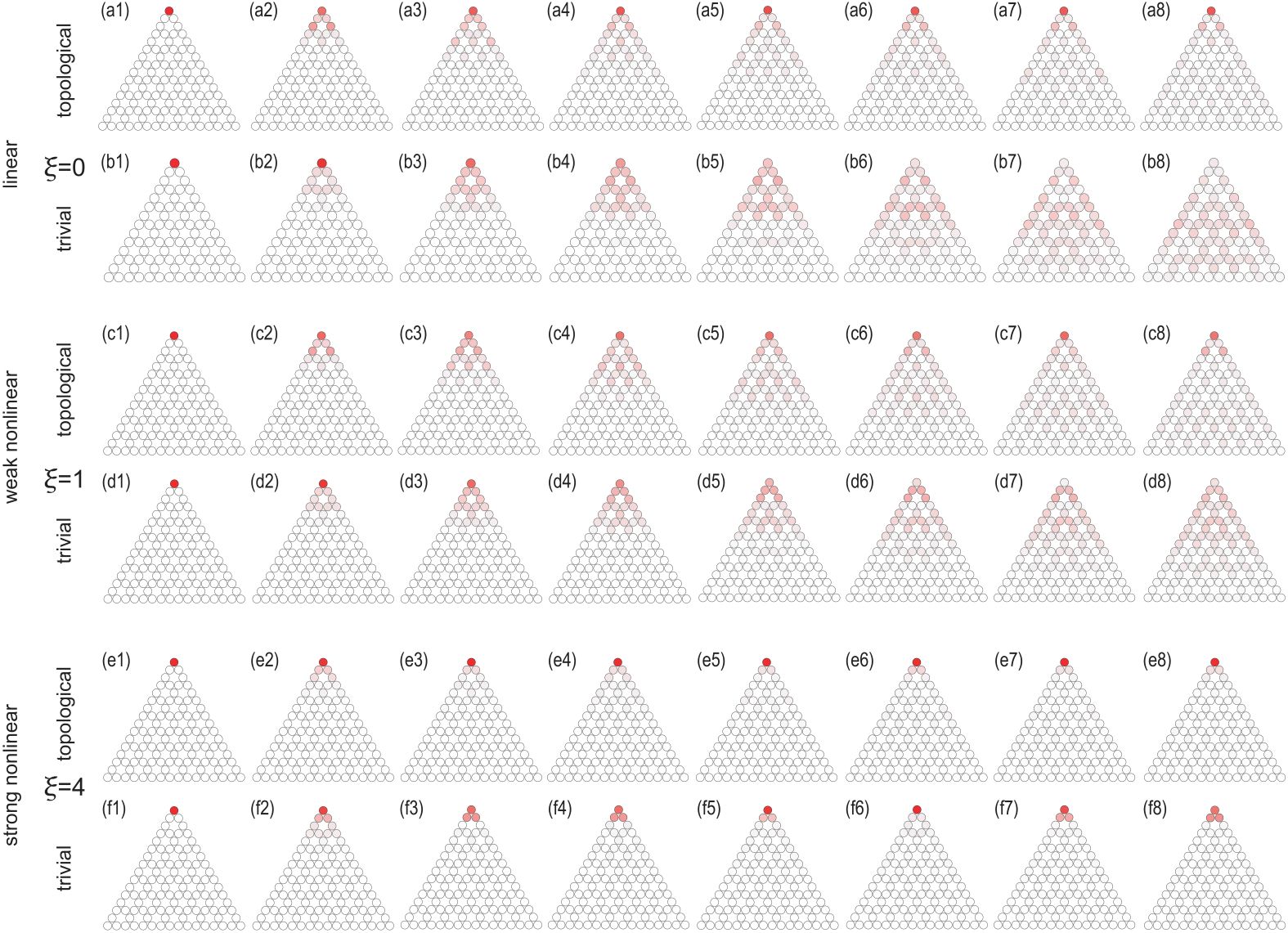}}
\caption{Time evolution of the spatial distribution of the amplitude $|%
\protect\psi _{n}|$ in the nonlinear breathing Kagome model. (a1)$\sim$(b8)
linear model with $\protect\xi =0$ for various time. (c1)$\sim$(d8) weak
nonlinear model with $\protect\xi =1$. (e1)$\sim$(f8) strong nonlinear model
with $\protect\xi =4$, where the system is in the nonlinearity-induced
localized phase. The color density indicates the amplitude $|\protect\psi %
_{n}|$ . We have set $\protect\lambda =-0.5$ for (a1)$\sim$(a8), (c1)$\sim$%
(c8) and (e1)$\sim$(e8), where the system is topological, while we have set $%
\protect\lambda =0.5$ for (b1)$\sim$(b8), (d1)$\sim$(d8) and (f1)$\sim$(f8),
where the system is trivial. }
\label{FigKagomeColor}
\end{figure*}

\subsection{Dimer limit}

\label{SecDimer}

Next, we study\ the dimer limit with $t_{B}=0$ as in Fig.\ref%
{FigKagomeIllust}(a3), where $\lambda =1$. The differential equations
are explicitly given by 
\begin{align}
i\frac{d\psi _{1}}{dt}& =t_{A}\left( \psi _{2}-\psi _{1}\right) -\xi
\left\vert \psi _{1}\right\vert ^{2}\psi _{1},  \label{dimer1} \\
i\frac{d\psi _{2}}{dt}& =t_{A}\left( \psi _{1}-\psi _{2}\right) -\xi
\left\vert \psi _{2}\right\vert ^{2}\psi _{2}.  \label{dimer2}
\end{align}%
We show a numerical solution of the time evolution of $\psi _{1}$ and $\psi
_{2}$ in Fig.\ref{FigDimerDynamics}.

In the linear model ($\xi =0$),\ they oscillate alternately without changing
their amplitudes,%
\begin{equation}
\psi _{1}=e^{-it_{A}t}\cos t_{A}t,\qquad \psi _{2}=ie^{-it_{A}t}\sin t_{A}t,
\end{equation}%
where the phases are different by $\pi $. Once the nonlinearity is
introduced, there appears an oscillation whose period is much longer than
the original period. The overall oscillation period becomes shorter as the
nonlinearity increases. It shows a complicated behavior for strong
nonlinearity\ as in Fig.\ref{FigDimerDynamics}.

In the nonlinear model ($\xi \neq 0$),\ there is an oscillatory behavior
with long and short periods in the dimer phase as in Fig.\ref%
{FigDimerDynamics}. This is easily seen by examining the Fourier component $%
\psi \left( \omega \right) $, where $\omega $\ is the frequency as\ in Fig.%
\ref{FigDimerDynamics}(b1)$\sim $(b4).\ There are two sharp peaks in $\psi
\left( \omega \right) $, which indicates that there are short-period and
long-period modes.

This oscillatory behavior may be understood as follows. By using an ansatz%
\begin{equation}
\psi _{2}=-\psi _{1},  \label{Psi12}
\end{equation}%
the equations (\ref{dimer1}) and (\ref{dimer2}) are summarized to one
equation%
\begin{equation}
i\frac{d\psi _{1}}{dt}=-2t_{A}\psi _{1}-\xi \left\vert \psi _{1}\right\vert
^{2}\psi _{1},
\end{equation}%
whose solution is given by%
\begin{equation}
\theta _{1}=\left( 2t_{A}+\xi r_{n}^{2}\right) t+c,  \label{DiAna}
\end{equation}%
and a constant $r_{1}$ with the polar expression (\ref{rt}). This explains a
short-period oscillation mode in Fig.\ref{FigDimerDynamics}.

However, the ansatz (\ref{Psi12}) for the analytical solution is not
compatible to the initial condition (\ref{IniCon}). Thus, we cannot apply
the analytic solution (\ref{DiAna}) for the quench dynamics, although the
ansatz (\ref{Psi12}) holds after enough time.\ In order to adjust the ansatz
(\ref{Psi12}) to the initial condition (\ref{IniCon}), the long-period
oscillation mode would appear.

These dimer oscillations give rise to the dimer phase in the vicinity of $%
\lambda =1$ in the phase diagram in Fig.\ref{FigSSHDiagram}(a3).

\begin{figure*}[t]
\centerline{\includegraphics[width=0.98\textwidth]{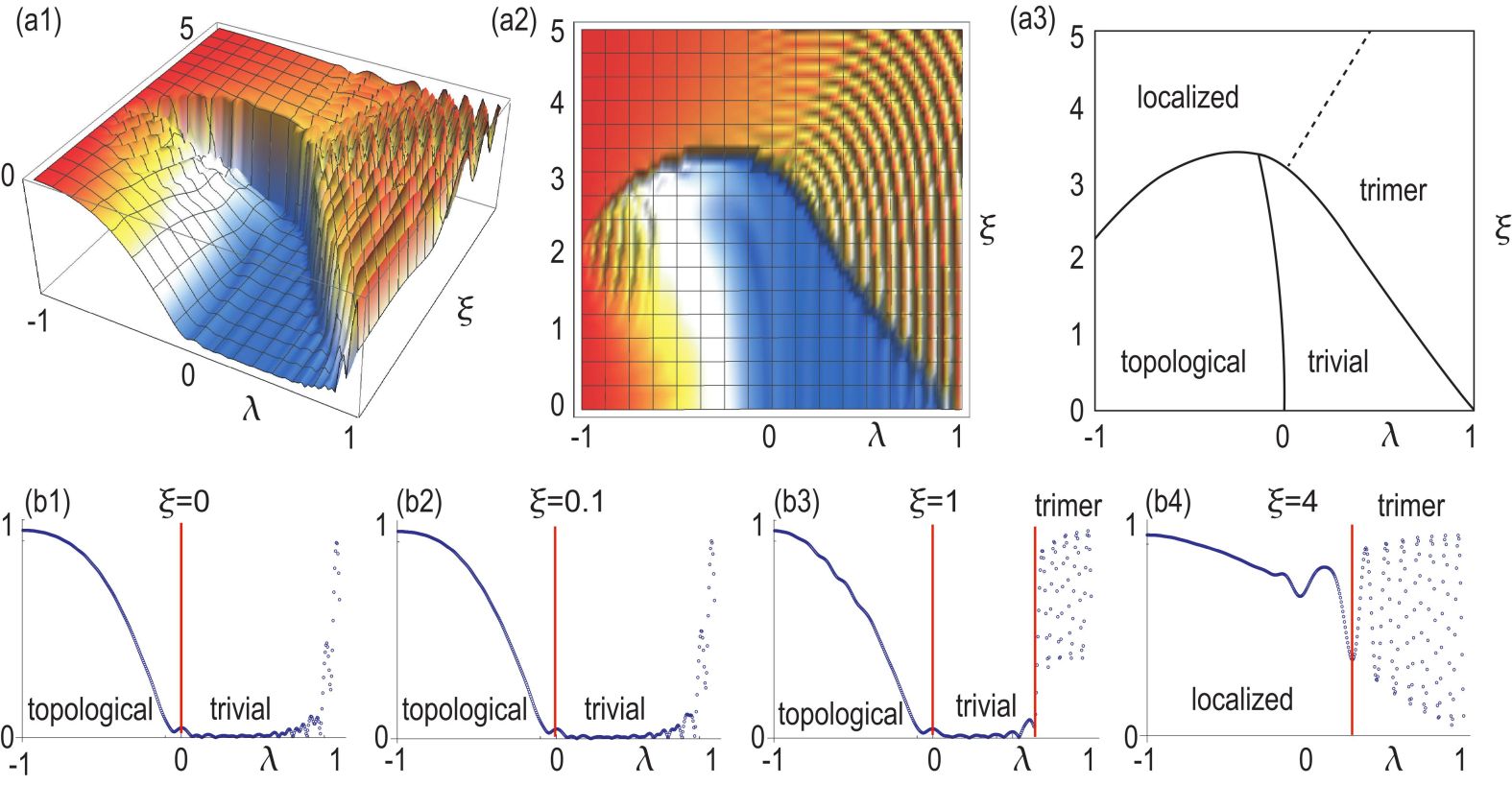}}
\caption{(a1)$\sim $(a3) Phase diagram of the nonlinear breathing Kagome
model. (a1) Bird's eye's view, (a2) top view, and (a3) schematic
illustration of the phase diagram. (b1)$\sim $(b4) Amplitude $|\protect\psi %
_{1}|$ as a function of $\protect\lambda $ for various $\protect\xi $. (b1) $%
\protect\xi =0$, (b2) $\protect\xi =0.1$, (b3) $\protect\xi =1$, and (b4) $%
\protect\xi =5$.}
\label{FigKagomeDiagram}
\end{figure*}

\subsection{Discrete nonlinear Schr\"{o}dinger equation}

\label{SecDNS}

The discrete nonlinear Schr\"{o}dinger equation (\ref{DNLS}) is a limit of
the nonlinear SSH model (\ref{EqA}) and (\ref{EqB}) by setting $\lambda =0$.
We show the time evolution of $\psi _{n}$ starting from the initial
condition (\ref{IniCon}), where the initial state is taken at the site $m$
in the bulk in Fig.\ref{FigLocalized}. For weak nonlinearity $\xi \lesssim 4$%
, the state rapidly spreads as shown in Fig.\ref{FigLocalized}(b2)$\sim $%
(b7). On the other hand, for strong nonlinearity $\xi \gtrsim 4$, the state
almost remains at the initial site $m$ as shown in Fig.\ref{FigLocalized}%
(b6) and (b7). We also show $\left\vert \psi _{1}\right\vert $ as a function
of $\xi $ in Fig.\ref{FigLocalized}(b1). It shows a drastic change at $\xi
\simeq 4$ for $\lambda =0$, which indicates the nonlinearity-induced
localization transition.

We have also shown the time evolution of $\psi _{n}$\ starting from the
initial condition (\ref{IniCon}) in the case of the nonlinear SSH model in
Fig\ref{FigLocalized}(a2)$\sim $(a7) for the topological phase with $\lambda
=-0.5$ and Fig\ref{FigLocalized}(c2)$\sim $(c7) for the trivial phase with $%
\lambda =0.5$. The nonlinearity-induced localization transition is found to
occur at $\xi \simeq 2$.

It is to be noted in Fig\ref{FigLocalized} that there is almost no
difference in the dynamics of $\left\vert \psi _{n}\right\vert $ between the
topological and trivial phases for $\lambda =\mp 0.5$. It dictates that we
cannot differentiate the topological and trivial phases by starting from a
site in the bulk. It is because the bulk state is almost identical between
the topological and trivial phases. The key difference is the presence
of topological edge or corner states in the topological phase.

As a result, the nonlinearity-induced localization occurs irrespective of
the dimerization $\lambda $. It is because the nonlinearity effect is
dominant in the dynamics for large $\xi $.

\section{Nonlinear second-order topological phases\label{SecKagome}}

\subsection{Model}

Recently, the nonlinear second-order topological phase has been studied in
photonics\cite{Kirch}. We proceed to study the case where the matrix $M_{nm}$
describes the breathing Kagome lattice, whose lattice structure is
illustrated in Fig.\ref{FigKagomeIllust}(b). The matrix $M$ in the momentum
space is given by\cite{EzawaKagome} 
\begin{equation}
M\left( \mathbf{k}\right) =-\left( 
\begin{array}{ccc}
0 & h_{12} & h_{13} \\ 
h_{12}^{\ast } & 0 & h_{23} \\ 
h_{13}^{\ast } & h_{23}^{\ast } & 0%
\end{array}%
\right) ,  \label{H3}
\end{equation}%
with 
\begin{align}
h_{12}& =t_{A}e^{i\left( k_{x}/2+\sqrt{3}k_{y}/2\right) }+t_{B}e^{-i\left(
k_{x}/2+\sqrt{3}k_{y}/2\right) }, \\
h_{23}& =t_{A}e^{i\left( k_{x}/2-\sqrt{3}k_{y}/2\right) }+t_{B}e^{i\left(
-k_{x}/2+\sqrt{3}k_{y}/2\right) }, \\
h_{13}& =t_{A}e^{ik_{x}}+t_{B}e^{-ik_{x}},
\end{align}%
where we have introduced two hopping parameters $t_{A}$ and $t_{B}$
corresponding to upward and downward triangles, as shown in Fig.\ref%
{FigKagomeIllust}(b).

\subsection{Topological number}

There are three mirror symmetries for the breathing Kagome lattice. They are
the mirror symmetries $M_{x}$ with respect to the $x$ axis, and $M_{\pm }$
with respect to the two lines obtained by rotating the $x$ axis by $\pm 2\pi
/3$. The polarization along the $x_{i}$ axis is the expectation value of the
position, 
\begin{equation}
p_{i}=\frac{1}{S}\int_{\text{BZ}}A_{i}d^{2}\mathbf{k},  \label{PolarP}
\end{equation}%
where $A_{i}=-i\left\langle \psi (\mathbf{k})\right\vert \partial
_{k_{i}}\left\vert \psi (\mathbf{k})\right\rangle $ is the Berry connection
with $x_{i}=x,y$, and $S=8\pi ^{2}/\sqrt{3}$ is the area of the Brillouin
zone; $\psi (\mathbf{k})$ the eigenfunction of $M\left( \mathbf{k}\right) $.

The topological number is defined by\cite{EzawaKagome} 
\begin{equation}
\Gamma =3\left( p_{x}^{2}+p_{y}^{2}\right) .
\end{equation}%
We obtain $\Gamma =0$ for $t_{A}/t_{B}<-1$ and $t_{A}/t_{B}>2$, which is the
trivial phase with no zero-mode corner states. On the other hand, we obtain $%
\Gamma =1$ for $-1<t_{a}/t_{b}<1/2$, which is the topological phase with the
emergence of three zero-mode corner states. Finally, $\Gamma $ is not
quantized for $1/2<t_{A}/t_{B}<2$, which is the metallic phase.

For $t_{A}=0$, three corner sites are perfectly decoupled\ from the bulk as
in Fig.\ref{FigKagomeIllust}(b1). On the other hand, for $t_{B}=0$, all
sites are trimerized as in Fig.\ref{FigKagomeIllust}(b3). We study a quench
dynamics starting from the initial condition (\ref{IniConA}), where the
state is perfectly localized at the top corner site. We note that we use the
tight-bind model although the continuum model is used in the previous work%
\cite{Kirch}, where the essential physics is identical.

\subsection{Phase diagram}

We show the phase diagram in Fig.\ref{FigKagomeDiagram}. There are four
phases in the nonlinear breathing Kagome model. The trimer phase appears
instead of the dimer phase characteristic to the nonlinear SSH model. The
trimer phase and the nonlinearity-induced localization phase are smoothly
connected, both of which are irrelevant to the topological number. The
difference is that there is an oscillatory behavior in the trimer phase but
not in the nonlinearity-induced localization phase. On the other hand, there
is a sharp transition between the trivial and nonlinearity-induced
localization phases. This is also the case for the transition between the
trivial and trimer phase. As in the case of the nonlinear SSH model, the
topological phase boundary between the topological and trivial phases is
almost unchanged for $0\leq \xi \lesssim 3$ as in Fig.\ref{FigKagomeDiagram}%
(a3).

We start with the study of the linear model ($\xi =0$). We show the spatial
distribution of the amplitude $|\psi _{n}|$ for various time in Fig.\ref%
{FigKagomeColor}(a1)$\sim $(a8) and Fig.\ref{FigKagomeColor}(b1)$\sim $(b8).
In the topological phase, the amplitude remains finite at the top corner
site. On the other hand, the amplitude rapidly spreads into the bulk and
disappears in the trivial phase.

The weak nonlinear regime ($\xi \simeq 0$) is analyzed just as in Sec.\ref%
{SecWeak}. Namely, the topological analysis based on the formula (\ref%
{TopoDynaA}) is valid as in the linear model. We have numerically confirmed
this observation in Fig.\ref{FigKagomeColor}(c1)$\sim $(c8). Indeed, we may
regard the system even with $\xi =1$ as the one in the weak nonlinear
regime. We show the spatial distribution for $\xi =1$ in Fig.\ref%
{FigKagomeColor}(c1)$\sim $(c8), which is almost identical to the one in the
linearized model ($\xi =0$) in Fig.\ref{FigKagomeColor}(a1)$\sim $(a8). This
is also the case for the trivial phase as shown in Fig.\ref{FigKagomeColor}%
(d1)$\sim $(d8). The amplitude at the top corner site after enough time is
shown in Fig.\ref{FigKagomeDiagram}(b1)$\sim $(b4). We note that the overall
feature is quite similar to the one in the nonlinear SSH model. On the other
hand, the state is localized in both of the topological and trivial phases
for $\xi =4$ in Fig.\ref{FigKagomeColor}(e1)$\sim $(e8) and Fig.\ref%
{FigKagomeColor}(f1)$\sim $(f8). It indicates that the state is in the
nonlinearity-induced localization phase.

\begin{figure}[t]
\centerline{\includegraphics[width=0.48\textwidth]{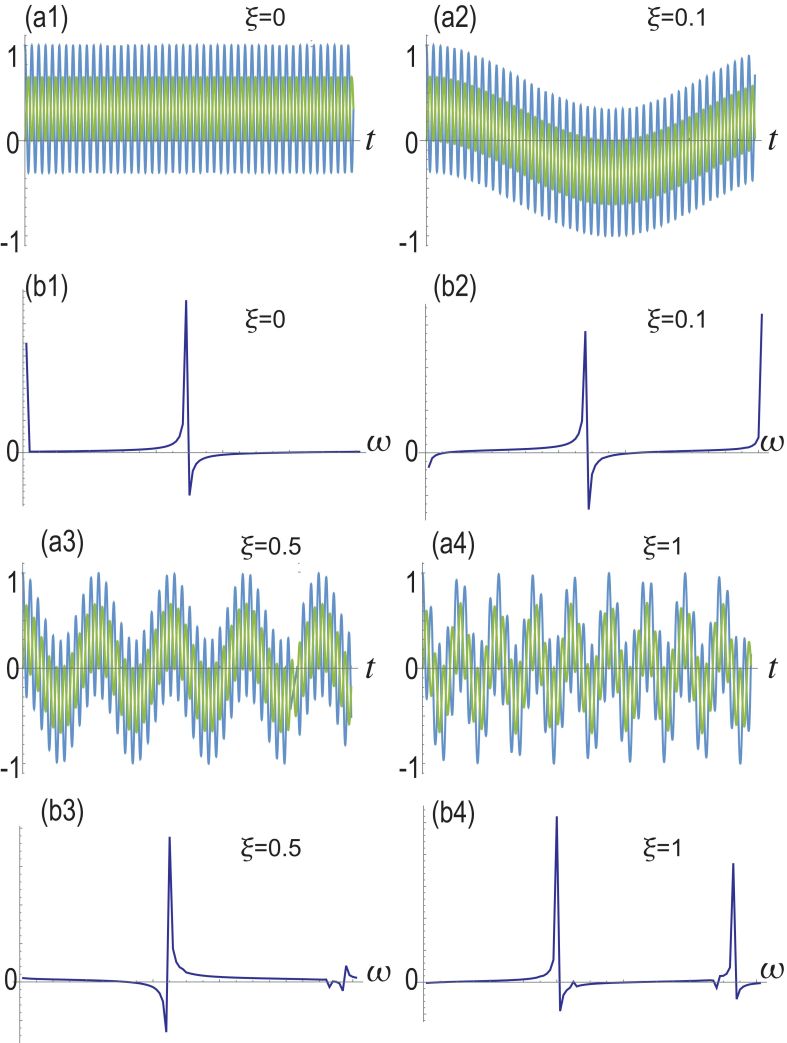}}
\caption{(a1)$\sim$(a4) Time evolution of Re[$\protect\psi_1(t)$] in the
nonlinear Schr\"{o}dinger model on the trimer described by Eqs.(\protect\ref%
{trimer1}), (\protect\ref{trimer2}) and (\protect\ref{trimer3}). (a1), (b1) $%
\protect\xi =0$; (a2), (b2) $\protect\xi =0.1$; (a3), (b3) $\protect\xi =0.5$%
; (a4), (b4) $\protect\xi =1$. (a1)$\sim$(a4) The vertical axis is Re[$%
\protect\psi_1(t)$] and the horizontal axis is time. (b1)$\sim$(b4) Fourier
component of Re[$\protect\psi_1(\protect\omega )$]. The horizontal axis is
the frequency $\protect\omega$, which is the Fourier component of the time $%
t $, while the vertical axis is Re[$\protect\psi_1(\protect\omega )$]. }
\label{FigTrimerDynamics}
\end{figure}

\subsection{Trimer limit}

Next, we study\ the trimer limit with $t_{B}=0$ as in Fig.\ref%
{FigKagomeIllust}(b3), where $\lambda =1$. The differential equations
are explicitly given by 
\begin{align}
i\frac{d\psi _{1}}{dt}& =\varepsilon \psi _{1}+t_{A}\left( \psi _{1}-\psi
_{2}\right) +t_{A}\left( \psi _{1}-\psi _{3}\right) -\xi \left\vert \psi
_{1}\right\vert ^{2}\psi _{1},  \label{trimer1} \\
i\frac{d\psi _{2}}{dt}& =\varepsilon \psi _{2}+t_{A}\left( \psi _{2}-\psi
_{1}\right) +t_{A}\left( \psi _{2}-\psi _{3}\right) -\xi \left\vert \psi
_{2}\right\vert ^{2}\psi _{2},  \label{trimer2} \\
i\frac{d\psi _{3}}{dt}& =\varepsilon \psi _{3}+t_{A}\left( \psi _{3}-\psi
_{1}\right) +t_{A}\left( \psi _{3}-\psi _{2}\right) -\xi \left\vert \psi
_{3}\right\vert ^{2}\psi _{3}.  \label{trimer3}
\end{align}%
Without loss of generality we may set $\psi _{2}=\psi _{3}$ and obtain%
\begin{align}
i\frac{d\psi _{1}}{dt}& =\varepsilon \psi _{1}+2t_{A}\left( \psi _{1}-\psi
_{2}\right) -\xi \left\vert \psi _{1}\right\vert ^{2}\psi _{1}, \\
i\frac{d\psi _{2}}{dt}& =\varepsilon \psi _{2}+t_{A}\left( \psi _{2}-\psi
_{1}\right) -\xi \left\vert \psi _{2}\right\vert ^{2}\psi _{2}.
\end{align}%
It is hard to solve\ these equations analytically except for the linear
model, where the solution is given by%
\begin{align}
\psi _{1}& =\frac{1}{3}\left( 1+2e^{-3it_{A}t}\right) , \\
\psi _{2}& =\psi _{3}=\frac{1}{3}\left( 1-e^{-3it_{A}t}\right) .
\end{align}%
Although this is modified smoothly as a function of $\xi $, it presents the
short-period oscillation.\ The long-period oscillation emerges once the
nonlinearity $\xi $ is introduced. We show the Fourier component $\psi
\left( \omega \right) $\ in Fig.\ref{FigTrimerDynamics}(b1)-(b4), where we
see two sharp peaks in $\psi \left( \omega \right) $ corresponding to the
short-period and long-period oscillations.

These trimer oscillations give rise to the trimer phase in the vicinity of $%
\lambda =1$\ in the phase diagram in Fig.\ref{FigKagomeDiagram}(a3).

\section{Conclusion}

The topological physics has been developed in linear systems such as
condensed matter systems, electric circuits and acoustic systems. The key
issue of the topological physics is the emergence of topological edge or
corner states in the topological phase, which has been firmly established by
various experimental observation. Now there are attempts generalize it to
nonlinear systems.

In the present work, we investigated the one-dimensional nonlinear SSH model
and the two-dimensional nonlinear breathing Kagome model. These models
contain the hopping term $\sum_{m}M_{nm}\psi _{m}$ and the nonlinear term $%
\xi \left\vert \psi _{n}\right\vert ^{2}\psi _{n}$. Dynamics is determined
as a result of the competition between these two terms. As far as the
hopping term is dominant, the topological dynamics is valid in these models.
On the other hand, when the nonlinear term is dominant, the
nonlinearity-induced localization phase emerges. There is another phenomenon
due to a cooperative effect of these two terms, which is the oscillation
mode in the dimer (trimer) limit of the nonlinear SSH (breathing Kagome)
model. We have studied these new phenomena analytically and numerically. Our
results are summarized in the phase diagrams in Fig.\ref{FigSSHDiagram} and
in Fig.\ref{FigKagomeDiagram}.

These results show that there are varieties of the effect of nonlinearity to
topological phases. It is an interesting problem to study various nonlinear
topological systems.

The author is very much grateful to N. Nagaosa for helpful discussions on
the subject. This work is supported by the Grants-in-Aid for Scientific
Research from MEXT KAKENHI (Grants No. JP17K05490 and No. JP18H03676). This
work is also supported by CREST, JST (JPMJCR16F1 and JPMJCR20T2).


\begin{thebibliography}{99}
\bibitem{Hasan} M. Z. Hasan and C. L. Kane, Rev. Mod. Phys. \textbf{82},
3045 (2010).

\bibitem{Qi} X.-L. Qi and S.-C. Zhang, Rev. Mod. Phys. \textbf{83}, 1057
(2011).

\bibitem{Fan} F. Zhang, C.L. Kane and E.J. Mele, Phys. Rev. Lett. \textbf{110%
}, 046404 (2013).

\bibitem{Science} W. A. Benalcazar, B. A. Bernevig, and T. L. Hughes,
10.1126/science.aah6442.

\bibitem{APS} F. Schindler, A. Cook, M. G. Vergniory, and T. Neupert, in APS
March Meeting (2017).

\bibitem{Peng} Y. Peng, Y. Bao, and F. von Oppen, Phys. Rev. B \textbf{95},
235143 (2017).

\bibitem{Lang} J. Langbehn, Y. Peng, L. Trifunovic, F. von Oppen, and P. W.
Brouwer, Phys. Rev. Lett. \textbf{119}, 246401 (2017).

\bibitem{Song} Z. Song, Z. Fang, and C. Fang, Phys. Rev. Lett. \textbf{119},
246402 (2017).

\bibitem{Bena} W. A. Benalcazar, B. A. Bernevig, and T. L. Hughes, Phys.
Rev. B \textbf{96}, 245115 (2017).

\bibitem{Schin} F. Schindler, A. M. Cook, M. G. Vergniory, Z. Wang, S. S. P.
Parkin, B. A. Bernevig, and T. Neupert, Science Advances 4, eaat0346 (2018).

\bibitem{FuRot} C. Fang, L. Fu, arXiv:1709.01929.

\bibitem{EzawaKagome} M. Ezawa, Phys. Rev. Lett. \textbf{120}, 026801 (2018).

\bibitem{Khalaf} E. Khalaf, H. C. Po, A. Vishwanath and H. Watanabe, Phys.
Rev. X \textbf{8}, 031070 (2018).

\bibitem{KhaniPhoto} A. B. Khanikaev, S. H. Mousavi, W.-K. Tse, M.
Kargarian, A. H. MacDonald, G. Shvets, Nature Materials \textbf{12}, 233
(2013).

\bibitem{Hafe2} M. Hafezi, E. Demler, M. Lukin, J. Taylor, Nature Physics 
\textbf{7}, 907 (2011).

\bibitem{Hafezi} M. Hafezi, S. Mittal, J. Fan, A. Migdall, J. Taylor, Nature
Photonics \textbf{7}, 1001 (2013).

\bibitem{WuHu} L.H. Wu and X. Hu, Phys. Rev. Lett. \textbf{114}, 223901
(2015).

\bibitem{TopoPhoto} L. Lu. J. D. Joannopoulos and M. Soljacic, Nature
Photonics \textbf{8}, 821 (2014).

\bibitem{Ozawa16} T. Ozawa, H. M. Price, N. Goldman, O. Zilberberg and I.
Carusotto Phys. Rev. A \textbf{93}, 043827 (2016).

\bibitem{Ley} D. Leykam and Y. D. Chong, Phys. Rev. Lett. \textbf{117},
143901 (2016).

\bibitem{KhaniSh} A. B. Khanikaev and G. Shvets, Nature Photonics \textbf{11}%
, 763 (2017).

\bibitem{Zhou} X. Zhou, Y. Wang, D. Leykam and Y. D. Chong, New J. Phys. 
\textbf{19}, 095002 (2017).

\bibitem{Jean} P. St-Jean, V. Goblot, E. Galopin, A. Lemaitre, T. Ozawa, L.
Le Gratiet, I. Sagnes, J. Bloch and A. Amo, Nature Photonics \textbf{11},
651 (2017).

\bibitem{Ota18} Y. Ota, R. Katsumi, K. Watanabe, S. Iwamoto and Y. Arakawa,
Communications Physics \textbf{1}, 86 (2018)

\bibitem{Ozawa} T. Ozawa, H. M. Price, A. Amo, N. Goldman, M. Hafezi, L. Lu,
M. C. Rechtsman, D. Schuster, J. Simon, O. Zilberberg and L. Carusotto, Rev.
Mod. Phys. \textbf{91}, 015006 (2019).

\bibitem{Ota19} Y. Ota, F. Liu, R. Katsumi, K. Watanabe, K. Wakabayashi, Y.
Arakawa and S. Iwamoto, Optica \textbf{6}, 786 (2019).

\bibitem{OzawaR} T. Ozawa and H. M. Price, Nature Reviews Physics \textbf{1}%
, 349 (2019).

\bibitem{Hassan} A. E. Hassan, F. K. Kunst, A. Moritz, G. Andler, E. J.
Bergholtz, M. Bourennane, Nature Photonics \textbf{13}, 697 (2019).

\bibitem{Ota} Y. Ota, K. Takata, T. Ozawa, A. Amo, Z. Jia, B. Kante, M.
Notomi, Y. Arakawa, S.i Iwamoto, Nanophotonics \textbf{9}, 547 (2020).

\bibitem{Li} M. Li, D. Zhirihin, D. Filonov, X. Ni, A. Slobozhanyuk, A. Alu
and A. B. Khanikaev, Nature Photonics \textbf{14}, 89 (2020).

\bibitem{Yoshimi} H. Yoshimi, T. Yamaguchi, Y. Ota, Y. Arakawa and S.
Iwamoto, Optics Letters \textbf{45}, 2648 (2020).

\bibitem{Kim} M. Kim, Z. Jacob and J. Rho, Light: Science and Applications 
\textbf{9}, 130 (2020).

\bibitem{Iwamoto21} S. Iwamoto, Y. Ota and Y. Arakawa, Optical Materials
Express \textbf{11}, 319 (2021).

\bibitem{Prodan} E. Prodan and C. Prodan, Phys. Rev. Lett. \textbf{103},
248101 (2009).

\bibitem{TopoAco} Z. Yang, F. Gao, X. Shi, X. Lin, Z. Gao, Y. Chong and B.
Zhang, Phys. Rev. Lett. \textbf{114}, 114301 (2015).

\bibitem{Berto} P. Wang, L. Lu and K. Bertoldi, Phys. Rev. Lett. \textbf{115}%
, 104302 (2015).

\bibitem{Xiao} M. Xiao, G. Ma, Z. Yang, P. Sheng, Z. Q. Zhang and C. T.
Chan, Nat. Phys. \textbf{11}, 240 (2015).

\bibitem{He} C. He, X. Ni, H. Ge, X.-C. Sun,Y.-B. Chen1 M.-H. Lu, X.-P. Liu,
L. Feng and Y.-F. Chen, Nature Physics \textbf{12}, 1124 (2016).

\bibitem{Abba} H. Abbaszadeh, A. Souslov, J. Paulose, H. Schomerus and V.
Vitelli, Phys. Rev. Lett. \textbf{119}, 195502 (2017).

\bibitem{Xue} H. Xue, Y. Yang, F. Gao, Y. Chong and B.Zhang, Nature
Materials \textbf{18}, 108 (2019).

\bibitem{Ni} X. Ni, M. Weiner, A. Alu and A. B. Khanikaev, Nature Materials 
\textbf{18}, 113 (2019).

\bibitem{Wei} M. Weiner, X. Ni, M. Li, A. Alu, A. B. Khanikaev, Science
Advances \textbf{6}, eaay4166 (2020).

\bibitem{Xue2} H. Xue, Y. Yang, G. Liu, F. Gao, Y. Chong and B. Zhang, Phys.
Rev. Lett. \textbf{122}, 244301 (2019).

\bibitem{Lubensky} C. L. Kane and T. C. Lubensky, Nature Phys. \textbf{10},
39 (2014).

\bibitem{Chen} B. Gin-ge Chen, N. Upadhyaya and V. Vitelli, PNAS \textbf{111}%
, 13004 (2014).

\bibitem{Nash} L. M. Nash, D. Kleckner, A. Read, V. Vitelli, A. M. Turner
and W. T. M. Irvine, PNAS \textbf{112}, 14495 (2015).

\bibitem{Paul} J. Paulose, A. S. Meeussen and V. Vitelli, PNAS \textbf{112},
7639 (2015).

\bibitem{Sus} R. Susstrunk, S. D. Huber, Science \textbf{349}, 47 (2015).

\bibitem{Sss} R. Susstrunk and S. D. Huber, Proc. Natl. Acad. Sci. USA 
\textbf{113}, E4767 (2016).

\bibitem{Huber} S. D. Huber, Nature Physics \textbf{12}, 621 (2016).

\bibitem{Mee} A. S. Meeussen, J. Paulose and V. Vitelli, Phys. Rev. X 
\textbf{6}, 041029 (2016).

\bibitem{Kariyado} T. Kariyado and Y. Hatsugai, Sci. Rep. \textbf{5}, 18107
(2016).

\bibitem{Hannay} T. Kariyado and Y. Hatsugai, J. Phys. Soc. Jpn. \textbf{85}%
, 043001 (2016).

\bibitem{Po} H. C. Po, Y. Bahri and A. Vishwanath, Phys. Rev. B \textbf{93},
205158 (2016).

\bibitem{Rock} D. Zeb Rocklin, Bryan Gin--ge Chen, Martin Falk, Vincenzo
Vitelli, and T.\thinspace C. Lubensky, Phys. Rev. Lett. \textbf{116}, 135503
(2016).

\bibitem{Takahashi} Y. Takahashi, T. Kariyado and Y. Hatsugai, New J. Phys. 
\textbf{19}, 035003 (2017).

\bibitem{Mat} K. H. Matlack, M. Serra-Garcia, A. Palermo, S. D. Huber and C.
Daraio, Nature Mat. \textbf{17}, 323 (2018).

\bibitem{Taka} Y. Takahashi, T. Kariyado and Y. Hatsugai, Phys. Rev. B 
\textbf{99}, 024102 (2019).

\bibitem{Ghatak} A. Ghatak, M. Brandenbourger, J. van Wezel and C. Coulais,
Proc. Natl. Ac. Sc. U.S.A. \textbf{117}, 29561 (2020).

\bibitem{Wakao} H. Wakao, T. Yoshida, H. Araki, T. Mizoguchi and Y.
Hatsugai, Phys. Rev. B \textbf{101}, 094107 (2020).

\bibitem{TECNature} S. Imhof, C. Berger, F. Bayer, J. Brehm, L. Molenkamp,
T. Kiessling, F. Schindler, C. H. Lee, M. Greiter, T. Neupert, R. Thomale,
Nat. Phys. \textbf{14}, 925 (2018).

\bibitem{ComPhys} C. H. Lee , S. Imhof, C. Berger, F. Bayer, J. Brehm, L. W.
Molenkamp, T. Kiessling and R. Thomale, Communications Physics, \textbf{1},
39 (2018).

\bibitem{Hel} T. Helbig, T. Hofmann, C. H. Lee, R. Thomale, S. Imhof, L. W.
Molenkamp and T. Kiessling, Phys. Rev. B \textbf{99}, 161114 (2019).

\bibitem{Lu} Y. Lu, N. Jia, L. Su, C. Owens, G. Juzeliunas, D. I. Schuster
and J. Simon, Phys. Rev. B \textbf{99}, 020302 (2019).

\bibitem{YLi} Y. Li, Y. Sun, W. Zhu, Z. Guo, J. Jiang, T. Kariyado, H. Chen
and X. Hu, Nat. Com. \textbf{9}, 4598 (2018).

\bibitem{EzawaTEC} M. Ezawa, Phys. Rev. B \textbf{98}, 201402(R) (2018).

\bibitem{Research} K. Luo, R. Yu and H. Weng, Research, ID 6793752. (2018).

\bibitem{Zhao} E. Zhao, Ann. Phys. \textbf{399}, 289 (2018).

\bibitem{EzawaLCR} M. Ezawa, Phys. Rev. B \textbf{99}, 201411(R) (2019).

\bibitem{EzawaSkin} M. Ezawa, Phys. Rev. B \textbf{99}, 121411(R) (2019).

\bibitem{Garcia} M. Serra-Garcia, R. Susstrunk and S. D. Huber, Phys. Rev. B 
\textbf{99}, 020304 (2019).

\bibitem{Hofmann} T. Hofmann, T. Helbig, C. H. Lee, M. Greiter, R. Thomale,
Phys. Rev. Lett. \textbf{122}, 247702 (2019).

\bibitem{EzawaMajo} M. Ezawa, Phys. Rev. B \textbf{100}, 045407 (2019).

\bibitem{Tjunc} M. Ezawa, Phys. Rev. B \textbf{102}, 075424 (2020).

\bibitem{Lee} C. H. Lee, T. Hofmann, T. Helbig, Y. Liu, X. Zhang, M. Greiter
and R. Thomale, Nature Communications \textbf{11}, 4385 (2020).

\bibitem{Kot} T. Kotwal, H. Ronellenfitsch, F. Moseley, A. Stegmaier, R.
Thomale, J. Dunkel, PNAS \textbf{118}, e2106411118 (2021).

\bibitem{Smi} D. Smirnova, D. Leykam, Y. Chong and Y. Kivshar, Applied
Physics Reviews \textbf{7}, 021306 (2020).

\bibitem{Kruk} S. Kruk, A. Poddubny, D. Smirnova, L. Wang, A. Slobozhanyuk,
A. Shorokhov, I. Kravchenko, B. Luther-Davies and Y. Kivshar, Nature
Nanotechnology \textbf{14}, 126 (2019).

\bibitem{MacZ} Lukas J. Maczewsky, Matthias Heinrich, Mark Kremer, Sergey K.
Ivanov, Max Ehrhardt, Franklin Martinez, Yaroslav V. Kartashov, Vladimir V.
Konotop, Lluis Torner, Dieter Bauer, Alexander Szameit, Science \textbf{370}%
, 701 (2010).

\bibitem{Zange} F. Zangeneh-Nejad and R. Fleury, Phys. Rev. Lett. \textbf{123%
}, 053902 (2019).

\bibitem{Kirch} M. S. Kirsch, Y. Zhang, M. Kremer, L. J. Maczewsky, S. K.
Ivanov, Y. V. Kartashov, L. Torner, D. Bauer, A. Szameit and M. Heinrich,
Nature Physics \textbf{17}, 995 (2021).

\bibitem{TopoToda} M. Ezawa, cond-mat/arXiv:2105.10851

\bibitem{MechaRot} M. Ezawa, arXiv:2108.09634.

\bibitem{Szameit} A. Szameit, D. Bl\"{o}er, J. Burghoff, T. Schreiber, T.
Pertsch, S. Nolte, A. T\"{u}nnermann and F. Lederer, Optics Express \textbf{%
13}, 10552 (2005).

\bibitem{Chris} D. N. Christodoulides, F. Lederer and Y. Silberberg, Nature 
\textbf{424}, 817 (2003)

\bibitem{Cai} D. Cai, A. R. Bishop and N. Gronbech-Jensen, Phys. Rev. Lett. 
\textbf{72}, 591 (1994).

\bibitem{Kev} P. G. Kevrekidis, K. O. Rasmussen and A. R. Bishop, Int. J.
Mod. Phys. B \textbf{15}, 2833 (2001).

\bibitem{Eil} J. C. Eilbeck, P. S. Lomdahl and A. C. Scott, Physica D 
\textbf{16}, 318, (1985).

\bibitem{Korab} N. Korabel and G. M. Zaslavsky, Physica A: Statistical
Mechanics and its Applications, \textbf{378}, 223 (2006).

\bibitem{Hadad} Y. Hadad, J. C. Soric, A. B. Khanikaev, and A. Al\`{u},
Nature Electronics \textbf{1}, 178 (2018).

\bibitem{Gor} M. A. Gorlach, A. P. Slobozhanyuk, Nanosystems: Physics,
Cheminstry, Mathematics, \textbf{8}, 695 (2017).

\bibitem{Tulo} T. Tuloup, R. W. Bomantara, C. H. Lee and J. Gong, Phys. Rev.
B \textbf{102}, 115411 (2020).

\end{thebibliography}
\end{document}